\documentclass[aps,prl,reprint,superscriptaddress,nofootinbib,showpacs,showkeys, 10pt]{revtex4-2}
\usepackage{graphicx}
\usepackage{xcolor}
\usepackage{caption}
\usepackage{subcaption}
\usepackage{float}
\usepackage{amsmath}
\usepackage{amsfonts}
\usepackage{physics}
\usepackage{comment}
\usepackage{amssymb}
\usepackage{pifont}
\usepackage{multirow}
\usepackage[most]{tcolorbox} % breakable boxes
\newcounter{protocol}
\usepackage[colorlinks=true,citecolor=blue,linkcolor=blue,urlcolor=blue]{hyperref}
\usepackage{booktabs}
\newtheorem{theorem}{Theorem}
\newtheorem{lemma}{Lemma}
\newtheorem{protocol}{Protocol}
\def\be{\begin{equation}}
\def\ee{\end{equation}}
\def\bea{\begin{eqnarray}}
\def\eea{\end{eqnarray}}
\def\ben{\begin{equation*}}
\def\een{\end{equation*}}
\def\bean{\begin{eqnarray*}}
\def\eean{\end{eqnarray*}}
\def\bma{\begin{mathletters}}
\def\ema{\end{mathletters}}
\def\bi{\begin{itemize}}
\def\ei{\end{itemize}}
\def\bd{\begin{description}}
\def\ed{\end{description}}

\begin{document}

\title{Device-Independent Anonymous Communication in Quantum Networks}

\author{Srijani Das}
\email{dassrijani33@gmail.com}
\email{srijani24_r@isical.ac.in}
\affiliation{Physics and Applied Mathematics Unit, Indian Statistical Institute, Kolkata 700108, India}
\author{Manasi Patra}
\email{manasipatra58@gmail.com}
\affiliation{Physics and Applied Mathematics Unit, Indian Statistical Institute, Kolkata 700108, India}
\author{Tuhin Paul}
\email{tuhin.paul96@gmail.com}
\affiliation{Physics and Applied Mathematics Unit, Indian Statistical Institute, Kolkata 700108, India}
\author{Anish Majumdar }
\email{anishmajumdar2020@gmail.com}
\affiliation{Computer and Communication Sciences Division, Indian Statistical Institute, Kolkata 700108, India}
\affiliation{Virtusa, Hyderabad, India}
\author{Ramij Rahaman}
\email{ramijrahaman@isical.ac.in}
\affiliation{Physics and Applied Mathematics Unit, Indian Statistical Institute, Kolkata 700108, India}

\begin{abstract}
Anonymity is a fundamental cryptographic primitive that hides the identities of both senders and receivers during message transmission over a network. Classical protocols cannot provide information-theoretic security for such task, and existing quantum approaches typically depend on classical subroutines and multiple private channels, thereby weakening their security in fully adversarial settings. In this work, we introduce the first fully quantum protocol for anonymous communication in realistic quantum networks with a device-independent security proof.

\end{abstract}

\maketitle
\section{Introduction}
The quantum internet \cite{kimble2008} envisions a global network that connects users through quantum entanglement, enabling communication that is fundamentally secure against eavesdropping \cite{Acin2007,RevModPhys.92.025002}. Unlike classical networks, it relies on quantum correlations to transmit information in a manner that makes any interception detectable. As a result, quantum networks support secure communication and coordination among distant parties and lay the foundation for communication systems with intrinsic guarantees of privacy and security.

In today's classical digital era, communication is instantaneous and far-reaching, yet it often leaves behind traceable footprints. Malicious adversary can exploit such information to identify the origin and content of messages, potentially leading to retaliation. In such a situation, ensuring anonymous communication \cite{chaum1988dining, HAOVeto2006} and preserving privacy is not merely desirable but essential, particularly in the case of distributed network scenario. While privacy protects the content of a message from being revealed, anonymity hides the identities of the sender and receiver.

In the classical setting, the requirement of pairwise secure private channels and an honest majority limits the security of any anonymous schemes, particularly when the parties are physically distant. To overcome these limitations, several quantum schemes have been proposed to enhance the security of anonymous communication. In particular, GHZ states \cite{christandl2005quantum,Brassard,RahamanKarVeto,Thalacker2021} and $W$ states \cite{lipinska2018anonymous,Gong2022} have been employed for the anonymous distribution of entanglement and the anonymous teleportation of quantum states \cite{christandl2005quantum,Bennettteleport}. However, a significant improvement was made by Unnikrishnan \emph{et al.} \cite{unnikrishnan2019anonymity} in 2019, who proposed a quantum protocol for anonymous communication and the distribution of EPR \cite{EPR} pairs (maximally entangle states) in networks. Subsequently, Yang \emph{et al.} \cite{yang2020examining} made some modification to protect the protocol against denial-of-service attacks, in which adversaries could corrupt the shared entanglement without being detected. More recently, important progress has been made in anonymous conference key generation by Grasselli \emph{et al.} \cite{Grasselli2019,Grasselli2022} and Hahn \emph{et al.} \cite{PRXQuantum.1.020325}.

Although the above mentioned protocols are claimed to be quantum, they in fact rely on classical subroutines such as the `\emph{parity}' and `\emph{logical OR}' protocols, which require an exponential number of pairwise secure private communication channels. Consequently, these protocols inherit security vulnerabilities similar to those of classical schemes. Additionally, these protocols lack any device-independent (DI) certification of the underlying quantum resources most notably the GHZ correlations, thereby falling short of providing security in the true sense. DI certification plays a crucial role in any secure quantum protocol, as it verifies the authenticity of the shared correlations solely from the observed input-output measurement statistics, without relying on any assumptions about the internal functioning or trustworthiness of the measurement devices involved.

In this regard, we introduce secure quantum protocols for the Parity and Logical OR problems that remove the need for classical subroutines such as pairwise private channels, honest-majority assumptions, or trusted quantum resources. By exploring these protocols, we develop scheme for anonymous EPR pair generation between a sender and a receiver, along with a collision-detection mechanism capable of identifying multiple simultaneous senders. Leveraging the resulting distributed entanglement, our framework enables fully anonymous teleportation of quantum states or classical information. Moreover, we incorporate DI certification of the shared GHZ resource, and we analyze the security of the overall protocol within the $\epsilon$-security framework for anonymous communication. 

\section{Quantum anonymous communication}
We begin by outlining the communication scenario. We consider a network of $n$ agents and, for simplicity, restrict our attention to the case of odd $n$; the even $n$ case is analogous, with one agent holding two qubits handles one qubit according to its assigned role in the protocol and the other qubit according to the role of an agent who is neither sender nor receiver. In our protocols, the GHZ state \(\ket{\psi_n^+}=\left(\ket{0}^{\otimes n}+\ket{1}^{\otimes n}\right)/\sqrt{2}\) serves as the fundamental resource and may be distributed by an untrusted source. Within this setting, each agent may behave either honestly or maliciously. Consequently, it is necessary to authenticate the shared states in a DI manner. Therefore, at the very outset of each protocol, the parties perform a self-testing (DI authenticity) test of the GHZ state on a randomly selected subset of the large number of copies of the shared correlations. If the selected correlations pass the self-testing procedure, the remaining correlations are certified to be genuine GHZ correlations. The self-testing authentication of GHZ correlations is described in the next section.

Let us begin with the Quantum Parity protocol, which enables the agents to determine the parity of their input strings in the network.

\begin{protocol}{\textbf{Parity}}\label{QP}
\end{protocol}
\textbf{Input:} $n$ agents share a GHZ state \(|\psi_n^+\rangle\), each holding an input bit $x_i$ between 0 or 1.\\

\noindent\textbf{Goal:} Determine parity of the inputs: \( y = \oplus_{i=1}^n x_i\).

\begin{enumerate}
 \item Each party applies phase flip operator $\sigma_z$ on their respective qubit if \(x_i = 1\), otherwise does nothing. Then the shared state remains $\ket{\psi_n^+}$ if the parity of the inputs is even otherwise the state reduces to \(\ket{\psi_n^-}=\left(\ket{0}^{\otimes n}-\ket{1}^{\otimes n}\right)/\sqrt{2}\).
 \item Since \(|\psi_n^+\rangle\) and \(|\psi_n^-\rangle\) are orthogonal, the agents can distinguish them deterministically by measuring their individual qubit in the Hadamard-basis and broadcast the measurement results $\pm 1$. Hereafter, we adopt the encoding $\pm 1 \mapsto \{0,1\}$ for measurement outcomes; consequently, the XOR of the measurement results yields the parity of the inputs.
\end{enumerate}

Next we present the Quantum Logical-OR protocol also known as Anonymous Veto which enables a group of agents to compute the logical OR of their inputs while preserving the anonymity of each individual input, ensuring that any veto remains untraceable to its source. In this protocol as well as throughout we use the same security
parameter $S$ for simplicity; however, this is not necessary. 

\begin{protocol}{\textbf{Logical-OR}}\label{QLOR}
\end{protocol}
\textbf{Input:} $n$ agents share \( S \) copies of \( |\psi_n^+\rangle \), and each agent \( i \) holds a private input bit \( x_i \).\\

\noindent\textbf{Goal:} Compute inputs' logical OR : \( V = \bigvee _{i=1}^n x_i \).
\begin{enumerate}
 \item For each copy of \( |\psi_n^+\rangle \), each agent \( i \) sets \( p_i =0\) if $x_i=0$; otherwise $p_i$ is chosen at random between $0$ or $1$. 
 \item All agents execute the \emph{Parity} Protocol \ref{QP} with inputs \( \{p_i\}_{i=1}^n \), obtaining the output
 \(
 y = \oplus_{i=1}^n p_i.
 \)
 \item Repeat steps $1$ and $2$ for all \( S \) copies. 
 If the parity output \( y \) is never \( 1 \) in any round, set \( V = 0 \). Otherwise, set \( V = 1 \).
\end{enumerate}

 Now we demonstrate the Quantum Notification Protocol, which allows a sender to anonymously notify a designated receiver.
\begin{protocol}{\textbf{Notification}}\label{QN}
\end{protocol}

\noindent\textbf{Input:} Sender selects the receiver $r$.\\

\noindent\textbf{Goal:} Sender anonymously notifies the receiver $r$.\\

The protocol is executed for each agent $i \in \{1,\dots,n\}$. 
For the \emph{round of agent $i$}, agents share $S$ copies of the 
$n$-qubit GHZ state $|\psi_n^+\rangle$. For each such round:

\begin{enumerate}
 \item Each agent $j \neq i$, the value $p_j$ is chosen as follows: If $i = r$ and agent $j$ is the sender, then $p_j$ chosen randomly from $\{0,1\}$. Otherwise, set $p_j = 0$.
 \item The agents execute the Parity protocol \ref{QP} with inputs
 $\{p_j\}_{j=1}^n$, with the modification that in the round agent $i$ withholds her measurement outcome while all
 other agents broadcast theirs. Agent $i$ locally computes
 \(y_i=\oplus_{j=1}^n p_j\).
 \item Steps 1 and 2 are repeated for all $S$ copies. 
 If the parity output $y_i$ is never equal to $1$ in any of the $S$ repetitions, 
 then $y_i$ is set to $0$.

 \item If agent $i$ obtains $y_i = 1$, then she identifies herself as the receiver.
\end{enumerate}

To confirm that the Notification Protocol \ref{QN} has correctly assigned the legitimate receiver, agents execute the Authentication Protocol \ref{Authentication} described in Annexure-I.

We now present fully quantum anonymous entanglement generation protocol that establishes an EPR pair between a sender and a receiver with perfect anonymity. The protocol operates in two modes: Verification and Entanglement generation. In the Verification mode, receiver assess only the correctness of the other agents’ actions; whereas in the other mode generate EPR pair between the sender and the receiver. 

Before executing the main protocol, the parties first run a \textit{Collision Detection} protocol (Protocol \ref{QEG}) to ensure that there is exactly one active sender. The \textit{Collision Detection} protocol is described in Annexure-I. The parties proceed with the \textit{Anonymous Entanglement Generation} protocol only if the collision detection output is $V=1$; otherwise, the entire scheme is aborted.

\begin{protocol}{\textbf{Anonymous Entanglement Generation}}\label{QEG}
\end{protocol}
\noindent\textbf{Goal:} Generate EPR pair anonymously between the sender ($s$) and the receiver ($r$).

\begin{enumerate}
 \item \textit{Notification: }The agents execute the Notification Protocol \ref{QN} to notify the intended receiver.

 \item Agents share \( |\psi_n^+\rangle \).

 \begin{enumerate}
 \item All agents \( j \notin \{s,r\} \) measure their qubits in the Hadamard basis and publicly announce their measurement outcomes \( a_j \).
 \item Sender anonymously flips \(S\) fair coins. If all outcomes are heads, she sets \(x=0\); otherwise, she sets \(x=1\). 

\noindent$\bullet$ If \( x = 0 \), sender initiates \emph{Entanglement generation} mode:
 \begin{enumerate}
 \item Sender chooses a uniformly random bit \( b \) and broadcasts it.
 \item Sender applies a \(Z\) gate to her qubit if \( b = 1 \);
 otherwise, she does nothing.
 \end{enumerate}

\noindent$\bullet$ If \( x = 1 \), sender go for \emph{Verification mode}. 
 \begin{enumerate}
 \item Sender measures her qubit in the Hadamard basis and reveals the measurement outcome \( a_s \).
 \end{enumerate}

 \item Receiver selects a uniformly random bit \(b'\) and broadcasts it.
 \end{enumerate}

 \item\textit{Mode notification to the receiver: } All agents execute the Quantum Parity Protocol \ref{QP} using a copy of the GHZ state, where the sender inputs \(x\), the receiver inputs an independent uniformly random bit \(x_r\), and all remaining agents input \(0\). Let the protocol output be \(y\). 

 \noindent$\bullet$ Receiver computes \(y \oplus x_r\). If \(y \oplus x_r = 0\), the receiver proceeds with entanglement generation; otherwise, the receiver enters the verification mode.
 \item \textit{Entanglement generation: } 
 If \(y \oplus x_r = 0\), the receiver applies a \(Z\) gate to the qubit associated with Step 2 whenever \(
 b \oplus \left( \oplus_{j \not\in\{s,r\}} a_j \right) = 1 
 \).

 \item \textit{Agents’ Actions Verification:} 
 If \( y\oplus x_r = 1 \), receiver measures her qubit of Step-2 in the Hadamard basis and obtains the outcome \( a_r \) (say), and computes \( y' = \left(a_s \oplus a_r\right) \oplus_{j \not\in \{s,r\}} a_j\). Verification is accepted if \( y' = 0 \). Agents return to Step-2 and repeat the protocol for many times. 

 \item \textit{Receiver notifies the success of the protocol to the sender: } Receiver sets $\tilde{x}_r=0$ if the outcome $y'=0$ occurs in almost all repetitions of the protocol; otherwise, the receiver sets $\tilde{x}_r=1$ and aborts the protocol. Agents then execute the Quantum Parity Protocol \ref{QP} using an additional GHZ state, where the receiver inputs $\tilde{x}_r$, sender inputs a uniformly random bit $\tilde{x}_s$, and all other agents input \(0\). Let $\tilde{y}$ denote the protocol output. If $\tilde{y}\oplus \tilde{x}_s=1$, sender abort the protocol; otherwise, the sender and the receiver successfully share an EPR pair.
 \end{enumerate}
 
Note that the outcomes of the Parity protocols in Steps 3, 5 and 6 are fully random to all agents other than the sender and the receiver. Thus, the agents except the sender and the receiver have no information about the mode selection in Step-3 and the abort decision in Step 5 and 6. This feature is completely absent in all the existing protocols.

\section{Self-testing of GHZ correlation}\label{Self-test}
In view of the GHZ paradox \cite{GHZ1}, we define the Bell-type operator \( \hat{\mathcal{O}}=\hat{\mathcal{O}}_0-\sum_{i=1}^n \hat{\mathcal{O}}_i ,\label{bell_op}\) where \(\hat{\mathcal{O}}_i = X_1 \cdots X_{i-1} Y_i Y_{i+1} X_{i+2} \cdots X_n\), \(\hat{\mathcal{O}}_0 = X_1 X_2 \cdots X_n\) with the convention $n+1\equiv1$ (mod n) and $ X_i=\sigma_x$, $ Y_i=\sigma_y$ are the Pauli matrices for $i=1,2\dots,n$. Lemma \ref{LR_bound} in Annexure-II shows that, under any local-realistic (LR) theory, this operator satisfies the bound $|\langle \mathcal{O}\rangle_{LR}|\leq (n-1)$. The following theorem establishes that the maximal quantum value-equal to the algebraic maximum $\langle \mathcal{O}\rangle=n+1$ can be achieved only by the state $\ket{\psi_n^+}$. 

\begin{theorem}\label{unique_GHZ}
The maximum (minimum) algebraic value of \( \langle \hat{\mathcal{O}}\rangle \) can only be achieved by $|\psi_n^{+(-)}\rangle$ for odd $n$.
\end{theorem}

\noindent\textit{Proof:} 
 Note that $ \langle {\Psi_n}^\pm | \hat{\mathcal{O}} |
{\Psi_n}^\pm \rangle=\pm(n+1)$. To establish uniqueness, let $\rho_{\pm}$ be any state satisfying $\langle\hat{\mathcal{O}}\rangle_{\rho_{\pm}}=\pm(n+1)$. Since $\hat{\mathcal{O}}$ is Hermitian, the state $\rho_\pm$ admits a decomposition in the orthonormal eigenbasis of $\hat{\mathcal{O}}$. Lemma \ref{non_deg} in Annexure-II ensures that the maximal (respectively minimal) eigenvalue of $\hat{\mathcal{O}}$ is non-degenerate, and Lemma \ref{lemma_2} in Annexure-II shows that all other eigenvalues lie at least $4k$ below (above) the maximal (minimal) eigenvalue. Therefore, the extremal expectation value can be achieved only by the corresponding eigenstate. Hence, $\rho_{\pm}=|\psi_n^{\pm}\rangle\langle\psi_n^\pm|$.

We now extend the proof to a device-independent scenario by using the following modified Jordan decomposition lemma \cite{masaness}.
\begin{lemma}
 Let \( X \) and \( Y \) be two Hermitian operators acting on a Hilbert space \( \mathcal{H} \), each having only eigenvalues \( \pm1 \). Then there exists a decomposition of \( \mathcal{H} \) into a direct sum of subspaces, \(\mathcal{H}=\oplus_{\gamma \in \Gamma} \mathcal{H}^\gamma \), where each subspace \(\mathcal{H}^\gamma \) has dimension at most two and both operators \( X \) and \( Y \) leave each subspace \(\mathcal{H}^\gamma \) invariant. 
 
 Consequently, the operators \( X \) and \( Y \) admit a simultaneous block-diagonal representation of the form \(X = \oplus_{\gamma \in \Gamma} X^\gamma\) and \(Y = \oplus_{\gamma \in \Gamma} Y^\gamma\), 
where \( X^\gamma \) and \( Y^\gamma \) act on the corresponding subspace \( \mathcal{H}^\gamma \).
\end{lemma}
Since \( X \) and \( Y \) may arise from experimental setups or untrusted devices, their effective action on each subspace \( \mathcal{H}^\gamma \) is given by \(
X^\gamma = \Pi^\gamma X \Pi^\gamma, \quad Y^\gamma = \Pi^\gamma Y \Pi^\gamma,\)
where \( \Pi^\gamma \) is the projector onto \( \mathcal{H}^\gamma \). Following the decomposition the state \( \rho \) and the observable $\hat{\mathcal{O}}$ take the form \( \rho =\oplus_{\mu_1\dots \mu_n} p_{\mu_1\dots \mu_n} \rho^{\mu_1\dots \mu_n}\) and \(\hat{\mathcal{O}} =\oplus_{\mu_1\dots \mu_n}\hat{\mathcal{O}}^{\mu_1\dots \mu_n}\), where \( \hat{\mathcal{O}}^{\mu_1\dots \mu_n} = \hat{\mathcal{O}}_0^{\mu_1\dots \mu_n} - \sum_{i=1}^n \hat{\mathcal{O}}_i^{\mu_1\dots \mu_n}\) with \(\hat{\mathcal{O}_0}^{\mu_1\dots \mu_n}=X^{\mu_1}\dots X^{\mu_n}\), and \(\hat{\mathcal{O}_i}^{\mu_1\dots \mu_n}=X^{\mu_1}\dots X^{\mu_{i-1}}Y^{\mu_i}Y^{\mu_{i+1}}X^{\mu_{i+2}}\dots X^{\mu_n}\). Thus, \(
 \langle \hat{\mathcal{O}} \rangle_\rho = \sum_{\mu_1\dots \mu_n} p_{\mu_1\dots \mu_n} \, \langle \hat{\mathcal{O}}^{\mu_1\dots \mu_n} \rangle_{\rho^{\mu_1\dots \mu_n}}\), where \( \langle \hat{\mathcal{O}}^{\mu_1\dots \mu_n} \rangle_{\rho^{\mu_1,\dots, \mu_n}} = \mathrm{Tr}\left( \rho^{\mu_1\dots \mu_n}\, \hat{\mathcal{O}}^{\mu_1\dots \mu_n} \right) \).

If \( \langle \hat{\mathcal{O}} \rangle_\rho = \pm (n+1 )\), then necessarily each \( \langle \hat{\mathcal{O}} \rangle^{\mu_1\dots \mu_n} = \pm (n+1 ) \), as \( \sum p_{\mu_1\dots \mu_n}=1\), which implies that each sub-state of the decomposed state must be a GHZ state $\ket{\psi_n^\pm}$ in view of Theorem \ref{unique_GHZ} i.e., \(
\rho^{\mu_1\dots \mu_n} = |\psi^\pm_n\rangle \langle \psi^\pm_n|\). Thus, the global state takes the form \(\rho = |\Psi\rangle\langle\Psi|\), where \(
|\Psi\rangle = \oplus_{\mu_1\dots \mu_n} \sqrt{p_{\mu_1\dots \mu_n}} \, |\psi_n^\pm\rangle^{\mu_1\dots \mu_n} \). 

Now suppose that each party possesses an ancillary qubit initialized in \( |0\rangle' \). Each party applies the local map \(|\tau\rangle^{\mu_j} \otimes |0\rangle_j' \mapsto |0\rangle^{\mu_j} \otimes |\tau\rangle_j'\) for \(\tau\in\{0,1\}\). 
After this local operation, the joint state evolves as \(
|\Psi\rangle \otimes |0\rangle'^{\otimes n} \longrightarrow |\xi\rangle \otimes |\psi^\pm_n\rangle'\), where \( |\psi^\pm_n\rangle' \) is the ideal GHZ state on the ancilla qubits, and \( |\xi\rangle \) is an uncorrelated junk state.

In a realistic setting, if the Bell violation observed
\(\langle \hat{\mathcal{O}} \rangle = (n+1) - \epsilon\) for some \(\epsilon > 0\),
then the fidelity deficit $\delta$ of the corresponding state is bounded by
\begin{equation*}
\frac{\epsilon}{2(n-1)} \leq \delta \leq \frac{\epsilon}{4},
\label{EA}
\end{equation*}
as established in Lemma \ref{EA} of Annexure-II.

\section{Security analysis}
\noindent Let $E_{\epsilon}$ denote the event that Protocol \ref{QEG} does not abort in this noise scenario. We then have the following theorem.
\begin{theorem}\label{SOM_1}
 For all $\epsilon>0$,
\[ Pr[E_\epsilon]\leq \frac{2^{-S}\left(1-\frac{\epsilon}{4(n-1)}\right )^{S-1}}{1-(1-2^{-S})\left(1-\frac{\epsilon}{4(n-1)}\right)^{2}}.\]
\end{theorem}

\noindent\textit{Proof sketch:} The probability that the protocol does not abort is governed by two independent mechanisms. First, it depends on the probability that the shared state is selected for entanglement generation rather than being tested, which is fixed by the public coin-flip procedure. Second, it depends on the probability that all verification tests in the preceding rounds are passed successfully. This latter probability is determined solely by the success probability of the Quantum Parity Protocol \ref{QP}, denoted by \( \Pr[P_\epsilon] \). In addition, the abort condition arising from the final parity verification step is also governed by the same success probability. Combining these contributions allows us to derive an upper bound on \( \Pr[E_\epsilon] \). The full proof is given in Annexure-III.

If there are at least $k$ honest agents, the probability that malicious agents can correctly identify the sender is bounded by the theorem 
\begin{theorem}\label{Guess_1} For all $\epsilon>0$
\[
\Pr[\text{guess}] \le \frac{1}{k} + \sqrt{\epsilon}.
\]
\end{theorem}

\noindent\textit{Proof sketch: } 
If the shared state has high fidelity with the ideal GHZ state $\ket{\Psi_n^+}$, then for any two agents \(i\) and \(j\) acting as the sender, the resulting states after the local phase-flip operation ($\sigma_z$) have large mutual fidelity.
also have high mutual fidelity. Using the fidelity-trace distance relation, we derive an upper bound on the trace distance between these resulting states. We then consider the optimal probability of distinguishing among $
k$ states in an ensemble. By combining the established relationship between the discrimination probability and the trace distance, we obtain the desired result. The detailed proof is given in Annexure-III.

\section{Conclusion}
In this work, we explored the framework of device-independent anonymous communication over quantum networks, where both security and anonymity are guaranteed without relying on the internal functioning of the devices used. By exploiting the GHZ paradox, we constructed a Bell-type inequality that enables the self-testing of GHZ correlations. The certified GHZ correlations allow us to develop several anonymous protocols - including Parity, Logical OR (Anonymous Veto), Collision Detection and as well as to establish EPR pairs between legitimate parties in a network. Our schemes are entirely quantum, fully device-independent, and do not rely on classical sub-protocols or require any private channels, which often undermine security in fully adversarial settings, as seen in existing approaches. We also provide a detailed security and robustness analysis under realistic noise conditions. Our scheme can be extended to generate anonymous GHZ correlations among a subset of legitimate parties, thereby enabling device-independent anonymous conference key generation. Overall, our results show that anonymity can be implemented directly within the principles of quantum mechanics, offering a scalable and practical route toward anonymous communication in the emerging quantum internet. 

\section*{Acknowledgment}
Anish Majumdar acknowledges the Indian Statistical Institute, Kolkata, for the opportunity to pursue his M.Tech. (CS) dissertation under the supervision of R. Rahaman, during which a part of this work was carried out.

\bibliography{ref}

\section{Annexure-I}

\begin{protocol}{\textbf{Collision Detection between senders}}
\end{protocol}
 \textbf{Input:} An agent selects the input bit $x_i=1$ if she wishes to act as a sender; otherwise, she selects $x_i=0$.\\ 
\noindent \textbf{Goal:} Determine whether exactly one sender is active.\\ 
\noindent Agents first use \textit{Veto A} to check whether any sender is active, and then use \textit{Veto B} to determine whether more than one sender is active.
\begin{enumerate}
 \item \textit{Veto A}: Detect if any sender is active.
 \begin{itemize}
 \item Each agent sets \( x_i \) according to their action.
 \item Run the \textit{Logical-OR Protocol} \ref{QLOR} on \( \{x_i\} \). Let the output be \( V_A \).
 \end{itemize}
\noindent In \textit{Veto A}, the \emph{Logical-OR} scheme allows each sender to detect the presence of any other active senders.
 \item \textit{Veto B}: Detect collisions, only if \( V_A = 1 \).
 \begin{itemize}
 \item Each agent sets $b_i=1$ if $x_i=1$ and detects the presence of other sender(s) in \textit{Veto A}; otherwise, $b_i=0$.
 \item Run the \textit{Logical-OR} Protocol \ref{QLOR} on \( \{b_i\} \). Let the output be \( V_B \).
 \end{itemize}

 \item Agents set the final output as follows:
 \[
 V =
 \begin{cases}
 0, & \text{if } V_A = 0 \quad \text{(no sender)} \\
 1, & \text{if } V_A = 1 \text{ and } V_B = 0 \quad \text{(single sender)} \\
 2, & \text{if } V_A = 1 \text{ and } V_B = 1 \quad \text{(collision)}
 \end{cases}
 \]
\end{enumerate}

\begin{protocol}{\textbf{Authentication of the Receiver}}\label{Authentication}
\end{protocol}

\noindent\textbf{Input:} The agents share $S$ copies of the GHZ state $\ket{\psi_n^+}$.

\noindent\textbf{Goal:} To authenticate the receiver's identity.

\begin{enumerate}
 \item For each of the $S$ copies of $\ket{\psi_n^+}$, the agents execute the Parity Protocol \ref{QP}, where every agent except the sender uses as input the output obtained in the corresponding round of the Notification Protocol \ref{QN}. The sender's input is always set to $0$.
 
 \item The sender compares the final parity outcome across all $S$ rounds with the input used by the receiver in the Notification Protocol.
 
 If the values do not match within the allowed tolerance, the sender aborts the protocol.
\end{enumerate}

\section{Annexure-II}

\subsection{Bell-type inequality for GHZ correlation}
To demonstrate a GHZ type paradox for the GHZ correlations \be | {\Psi_n}^\pm \rangle = \frac{1}{\sqrt{2}} ( | 0 \rangle^{\otimes n} \pm | 1 \rangle^{\otimes n} ),\ee let us define the following $n+1$ observables for an odd $n$
\begin{equation}\label{ghz_op}
\begin{split}
 \hat{\mathcal{O}}_0 &= X_1 X_2 \cdots X_n, \\
 \hat{\mathcal{O}}_i &= X_1 \cdots X_{i-1} Y_i Y_{i+1} X_{i+2} \cdots X_n,  i =1,\dots, n,
\end{split}
\end{equation}
where $ n+1\equiv1$ (mod n) and $X_i=\sigma_x$, $Y_i=\sigma_y$ are the standard Pauli operators. One can easily verify that \( | {\Psi_n}^\pm \rangle \)
satisfy the following eigenvalue relations with the observables (\ref{ghz_op}):
\begin{equation}
\begin{split}
 \hat{\mathcal{O}}_0 | {\Psi_n}^\pm \rangle &= \pm | {\Psi_n}^\pm \rangle,\\
\hat{\mathcal{O}}_i | {\Psi_n}^\pm \rangle &= \mp | {\Psi_n}^\pm \rangle.
\end{split} 
\label{ghz_ev}
\end{equation}

This set of conditions cannot be satisfied by any local-realistic (LR) theory. The contradiction arises immediately when one attempts to assign predetermined LR values \( x_i, y_i \in \{ \pm 1 \} \) to each local observable \(X_i, Y_i\). Under such an assignment, the eigenvalue relations in (\ref{ghz_ev}) imply \begin{equation}
 \begin{split} 
 v(\hat{\mathcal{O}}_0) &= \prod_{i=1}^{n} x_i = \pm 1,\\
v(\hat{\mathcal{O}}_i) &= x_1 \cdots x_{i-1}y_iy_{i+1}x_{i+2} \cdots x_n = \mp 1. \end{split}
 \label{lrvalue}
\end{equation}
Now consider the product of all \( n+1 \) value assignments:
\begin{align}
\prod_{i=0}^n v(\hat{\mathcal{O}}_i)&= \prod_{i=0}^n (x_i)^{n-1} \cdot (y_i)^2 \nonumber\\&= (\pm 1)^{n-1} \cdot (+1) = +1, \label{value}\end{align}
since \( (y_i)^2 = +1 \) for all \( i \) and $n-1$ is even.

On the other hand, from (\ref{lrvalue})we have, \begin{equation*}
 \prod_{i=0}^n v(\hat{\mathcal{O}}_i) = (\pm 1) \cdot (\mp 1)^n = -1 \quad (\because  n \text{ is odd}).\label{cvalue}
\end{equation*}
This contradicts the value obtained in (\ref{value}). 

Let us now construct the following Bell-type operator 
\begin{align}
 \hat{\mathcal{O}}&=\hat{\mathcal{O}}_0-\sum_{i=1}^n \hat{\mathcal{O}}_i.
 \label{operator_O}
\end{align}
It is straightforward to verify that \begin{equation}
\label{eq:On-exp}
\langle \Psi_n^\pm | \hat{\mathcal{O}} | \Psi_n^\pm \rangle
= \pm (n+1),
\end{equation}
 which also corresponds to the algebraic optimal of \(\langle \hat{\mathcal{O}}\rangle \).
\begin{lemma}\label{LR_bound}
In any LR theory,
\[
{-(n-1) \leq \langle \hat{\mathcal{O}} \rangle_{LR} \leq (n-1)}.
\]
\end{lemma}

\noindent\textit{Proof. }
The possible algebraic values of $v(\hat{\mathcal{O}})$ lie in
\[
\{-(n+1),\, -(n-1),\, \dotsc,\, (n-1),\, (n+1)\}.
\]

However, as shown earlier (using equations (\ref{ghz_ev}) and (\ref{lrvalue})), not all such assignments are compatible with a LR model. In particular, any assignment that yields
\[
v(\hat{\mathcal{O}})=\pm (n+1) \quad \text{or} \quad v(\hat{\mathcal{O}})=\pm n
\]
violates the product constraint derived previously and therefore cannot arise in an LR theory. 
Thus, the only consistent LR assignments are those satisfying
\[
{|\langle \hat{\mathcal{O}} \rangle_{LR}| \leq (n-1)}.
\]

\begin{lemma}\label{non_deg}
 The optimal eigenvalues \(\pm(n+1)\) of $\mathcal{\hat{O}}$ are non-degenerative.
\end{lemma}
\noindent\textit{Proof.} 
To prove the lemma, it is sufficient to show that the eigenvectors corresponding to the optimal eigenvalues are non-degenerate, since $\mathcal{\hat{O}}$ is hermitian. Consider a general $n$-qubit pure state 
\begin{align}\label{eigenvector_g}
 |\psi\rangle=\sum_{b_1,\dots,b_n\in\{0,1\}} C_{b_1,\dots,b_n}|b_1\dots b_n\rangle,
\end{align}
where \(\displaystyle \sum_{b_1,\dots,b_n\in\{0,1\}}|C_{b_1,\dots,b_n}|^2=1\). Now,
\begin{align}\label{eigen_operator}
\hat{\mathcal{O}} |\psi\rangle &= \sum_{b_1, \dots, b_n\in\{0,1\}} C_{b_1 \dots b_n} \left[
1 + \sum_{i=1}^{n} (-1)^{b_i \oplus b_{i+1}}
\right] |\bar{b}_1 \dots \bar{b}_n\rangle,
\end{align}
\normalsize
where \(\bar{b}_i=b_i \oplus 1\). For convenience, define the coefficient
\begin{align}
 \label{O_value}
 \mathcal{O=} 1 + \sum_{i=1}^{n} (-1)^{b_i \oplus b_{i+1}}.
\end{align} 
The operator $\hat{\mathcal{O}}$ attains its maximum value $(n+1)$ only when every term in the expression \(\sum_{i=1}^{n} (-1)^{b_i \oplus b_{i+1}}\) is equal to $+1$. This occurs precisely when all bits are identical, i.e., $b_i=0$ for all $i$ or $b_i=1$ for all $i$. Hence, only the computational basis states $\ket{0}^{\otimes n}$ and $\ket{1}^{\otimes n}$ contribute. To satisfy the eigenvalue equation \begin{align}
 \hat{\mathcal{O}}|\psi\rangle &=\mathcal{O}|\psi\rangle,
\end{align} the corresponding coefficients must obey $C_{00\dots 0} =C_{11\dots 1}$. 
Thus, the only possibility (up to global phase) of $\ket{\psi}$ will be \begin{align*} 
|\psi\rangle=\frac{1}{\sqrt{2}}(|00\dots 0\rangle+|11\dots 1\rangle)=|\psi^{+}_n\rangle. 
\end{align*} 
This shows that the eigenvector corresponding to the largest eigenvalue $(n+1)$ is non-degenerate. Similarly, one can show that the eigenvector $\ket{\Psi_n^-}$, which corresponds to the smallest eigenvalue of $\hat{\mathcal{O}}$ is also non-degenerate.
\begin{lemma}\label{lemma_2}
 The eigenvalues of the operator \( \hat{\mathcal{O}} \) are of the form
\(\pm\left[( n+1) - 4k \right]\), with \(k = 0, 1, \dots, \left\lfloor \dfrac{n+1}{4} \right\rfloor\).
\end{lemma}
\textit{Proof}---
Equation (\ref{eigen_operator}) shows that every eigenvector of \( \hat{\mathcal{O}} \) must be of the form \begin{align}
 \ket{\psi}=C_{b_1\dots b_n}\ket{b_1\dots b_n}+C_{\bar{b}_1\dots\bar{b}_n}\ket{\bar{b}_1\dots\bar{b}_n},
\end{align} with $C_{b_1\dots b_n} =\pm C_{\bar{b}_1\dots \bar{b}_n}$ and \(|C_{b_1\dots b_n}|^2+|C_{\bar{b}_1\dots \bar{b}_n}|^2=1\).
Define $f$ as the number of consecutive anti-correlated bit pairs (i.e. $ b_i\neq b_{i+1}$), in the string of $\ket{\psi}$.
From equation (\ref{O_value}), each anti-correlated pair contributes a factor of $(-1)$, so exactly $f$ terms contribute $(-1)$. Therefore, $\mathcal{O}=1+(n-f)-f=n+1-2f$.\\ 
 Moreover, the bit string is cyclic and $n$ is odd, the number of anti-correlations \( f \) must be even, so we set $f=2k$. Thus the eigenvalues of \( \hat{\mathcal{O}} \) take the form:
\[
\pm\left( (n+1) - 4k \right), \quad \text{for } k = 0, 1, 2, \dots, \left\lfloor \frac{n+1}{4} \right\rfloor.
\]

\begin{lemma}
Let $\delta$ denote the fidelity deficit of an imperfect state $\tilde{\rho}$ relative to the ideal state $\ket{\psi_n^+}$, which exhibits a Bell violation $\langle \hat{\mathcal{O}} \rangle_{\tilde{\rho}} = (n+1)-\epsilon$, for some $\epsilon>0$. Then the fidelity deficit satisfies, 

\[\frac{\epsilon}{2(n-1)}\leq\delta\leq\frac{\epsilon}{4}.\] \label{EA}
\end{lemma}
\textit{Proof. } We can express the imperfect state $\tilde{\rho}$ as
\[
\tilde{\rho} = (1 - \delta) |\psi_n^+\rangle\langle\psi_n^+| + \delta \, \sigma,
\]
where \(\delta = 1 - \Tr \left( \tilde{\rho} \, |\psi_n^+\rangle\langle \psi_n^+| \right)\) and \(\displaystyle \text{Tr}\left[|\Psi_n^+\rangle\langle \Psi_n^+|\sigma\right]=0\).
The observable $\hat{\mathcal{O}}$ attains its maximal value $n+1$ on the state
$|\psi_n^+\rangle$ [ see Eq. \eqref{eq:On-exp}] and for any state orthogonal to
$|\psi_n^+\rangle$ its expectation value lies in the interval
$[-(n-3),(n-3)]$.
 Therefore,
\begin{align*}
\Tr\left[ \tilde{\rho}\hat{\mathcal{O}} \right]
&= (n+1)-\epsilon=(1 - \delta)(n+1) + \delta\alpha,
\end{align*}
$\text{ where } \alpha \in [-(n-3), (n-3)]$.
Thus, \begin{align*}
 \frac{\epsilon}{2(n-1)}\leq\delta\leq\frac{\epsilon}{4}.\end{align*}
 
\section*{Annexure-III}
\begin{lemma}\label{SOP}
 Let $P_{\epsilon}$ denote the event that the Parity Protocol \ref{QP} does not abort when an imperfect state with Bell violation $\langle \hat{\mathcal{O}} \rangle_{\tilde{\rho}} = (n+1)-\epsilon$, for some $\epsilon>0$, is used as the resource. Then 
\[\left(1-\frac{\epsilon}{4}\right)\leq Pr[P_\epsilon]\leq \left(1-\frac{\epsilon}{4(n-1)}\right).\]
\end{lemma}
\noindent \textit{Proof. }For any $n$-qubit state that can be expressed as \(
\tilde{\rho} = (1-\delta)\ket{\psi_n^+}\bra{\psi_n^+} + \delta\,\sigma
\),
the probability $\Pr[P_\epsilon]$ of obtaining an even-parity outcome from $\tilde{\rho}$ in a single run is given by:
\begin{align*}
 (1-\delta)\leq Pr[P_\epsilon]\leq 
 \left( 1-\frac{\delta}{2}\right).
\end{align*} 
Using Lemma \ref{EA}, we have
\begin{align*} \left(1-\epsilon/4\right)\leq Pr[P_\epsilon]\leq\left(1-\frac{\epsilon}{4(n-1)}\right).\end{align*}

\subsection{Proof of Theorem 2}
\setcounter{theorem}{1}
\begin{theorem}\label{SOM_1}
Let $E_{\epsilon}$ denote the event that Protocol \ref{QEG} does not abort when the shared resource is an imperfect state exhibiting a Bell violation $\langle \hat{\mathcal{O}} \rangle = (n+1)-\epsilon$, for some $\epsilon>0$. Then the probability of this event satisfies
\[ Pr[E_\epsilon]\leq \frac{2^{-S}\left(1-\frac{\epsilon}{4(n-1)}\right )^{S-1}}{1-(1-2^{-S})\left[1-\frac{\epsilon}{4(n-1)}\right]^{2}}.\]

\end{theorem}
\noindent\textit{Proof. }Firstly, the probability that the receiver is correctly identified during the notification stage is $(1-2^{-S})\,Pr[P_\epsilon]^{S}$. Next, consider the probability that the state is used to generate an EPR pair in round $\ell$. For this to occur, the sender must obtain all $S$ coin flips as heads ($x=0$) in Step-3, which happens with probability $2^{-S}$. The probability that the state is tested in each of the preceding $(\ell-1)$ rounds is $(1 - 2^{-S})$. In Step-5, the probability of obtaining the correct outcome $y'=0$ is $Pr[P_\epsilon]$. Thus, the success probability for each tested round is $Pr[P_\epsilon]$. Finally, another Parity Protocol \ref{QP} is executed to convey the receiver's decision to the sender regarding whether the protocol should abort at Step-6. Combining all these, the overall probability of event $E_\epsilon$ occurring in the $\ell$-th repetition of the protocol is \begin{align*}
 Pr[E_\epsilon^\ell]& =2^{-S}(1-2^{-S})^{\ell}{Pr[P_\epsilon]}^{S+2\ell-1}\\ &\leq 2^{-S}(1-2^{-S})^{\ell}\left[1-\frac{\epsilon}{4(n-1)}\right]^{S+2\ell-1}.
\end{align*}
To obtain an upper bound on the total probability, we integrate over $\ell$:
 \begin{align*}
 Pr[E_\epsilon]&\leq 2^{-S}\int_{0}^{\infty}
(1-2^{-S})^{\ell}\left[1-\frac{\epsilon}{4(n-1)}\right]^{2\ell+(S-1)}\,dl
\\ &\leq -\frac{2^{-S}\left[1-\frac{\epsilon}{4(n-1)}\right]^{S-1}}{\ln\left[(1-2^{-S})\left[1-\frac{\epsilon}{4(n-1)}\right]^{2}\right]}\\&\leq \frac{2^{-S}\left[1-\frac{\epsilon}{4(n-1)}\right ]^{S-1}}{1-\left[(1-2^{-S})\left[1-\frac{\epsilon}{4(n-1)}\right]^{2}\right]}.
 \end{align*}

\subsection{Proof of Theorem 3}
\begin{theorem}\label{Guess_1}
If the Protocol \ref{QEG} is executed with at least $k$ honest agents and the shared resource is an imperfect state exhibiting a Bell violation $\langle \hat{\mathcal{O}} \rangle = (n+1)-\epsilon$, for some $\epsilon>0$. Then, the probability that the malicious agents correctly identify the sender is upper bounded by \[
\Pr[\text{guess}] \le \frac{1}{k} + \sqrt{\epsilon}.
\]
 \end{theorem}
\textit{Proof. }
Any mixed state can be expressed as a probabilistic mixture of pure states, and the overall guessing probability is therefore a convex combination of the guessing probabilities corresponding to the constituent pure states. Consequently, it suffices to restrict our analysis to pure-state strategies.

If there are $k$ honest agents, any one of whom may act as the sender. Our goal is thus to determine the maximum success probability of a measurement that distinguishes among the $k$ states arising from different agents applying a local phase-flip operation $\sigma_z$.

Suppose the agents share a state of the form \[\ket{\psi}=\sqrt{(1-\delta)}\ket{\psi_n^+}+\delta_1\ket{\psi_n^-}+\sum_{i=2}^{2^n-1}\delta_i\ket{\phi_n^{(i)}}.\]
For each possible sender applying $\sigma_z$ to their qubit, the resulting states satisfy the following bound on fidelity %and trace distance:
 \begin{align*}
 \mathcal{F}( |\psi_i \rangle, |\psi_j\rangle)&= |\langle\psi_i|\psi_j\rangle|\geq(1-2\delta)\\
 \end{align*} 
Utilizing the following relation between trace distance $\mathcal{D}(\ket{\psi_i},\ket{\psi_j})$ and fidelity $\mathcal{F}$,\[ \mathcal{D}(\ket{\psi_i},\ket{\psi_j})\leq \sqrt{1-{[\mathcal{F}( |\psi_i \rangle, |\psi_j\rangle)]}^2},\] we have the bound on trace distance as, \[\mathcal{D}(\ket{\psi_i},\ket{\psi_j})\leq2 \sqrt{\delta}.\]
Fixing an arbitrary reference state $\ket{\psi_j}$, the remaining states lie within trace distance $2\sqrt{\delta}$ of $\ket{\psi_j}$. For any POVM $\{\Pi_i\}$, this implies
\[\Tr(\Pi_i \ketbra{\psi_i}{\psi_i})-\Tr(\Pi_i \ketbra{\psi_j}{\psi_j})\leq \mathcal{D}(|\psi_{i}\rangle,\ket{\psi_j})\leq 2\sqrt{\delta}.\] 

The success probability for discriminating $k$ states in an ensemble $\{(p_i,\ketbra{\psi_i}{\psi_i})\}_{i=1}^k$, where each of the $k$ agents has an equiprobable chance of becoming the Sender (i.e., $p_i=\frac{1}{k}$), is given by
\begin{align*}
Pr[\text{guess}]
&=\sum_{i=1}^k(1/k)\Tr(\Pi_i\ketbra{\psi_i}{\psi_i})\\&\leq\frac{1}{k}\sum_{i=1}^k\left[\Tr(\Pi_i\ketbra{\psi_j}{\psi_j})+2\sqrt{\delta}\right]\\
&=\frac{1}{k}\Tr\left(\ketbra{\psi_j}{\psi_j}\right)+2\sqrt{\delta}\\&=\frac{1}{k}+2\sqrt{\delta}\\
&\leq\frac{1}{k}+\sqrt{\epsilon} [\text{By Lemma \ref{EA}}].
\end{align*}

\end{document}